\documentclass[aps,article,twocolumn,superscriptaddress]{revtex4}

\usepackage{amsmath}
\usepackage{amssymb}
\usepackage{graphicx}
\usepackage{xcolor}
\usepackage{epstopdf, epsfig}
\usepackage[utf8]{inputenc}
\usepackage{graphicx}
\usepackage{mathtools}
\usepackage{ulem}

\begin{document}
\title{Key parameters for surface plasma wave excitation in the ultra-high intensity regime}

\author{S. Marini} %\email{samuel.marini@polytechnique.edu}
\affiliation{LSI, CEA/DRF/IRAMIS, CNRS, \'Ecole Polytechnique, Institut Polytechnique de Paris,  F-91120 Palaiseau, France.}
\affiliation{LULI, Sorbonne Universit\'e, CEA, CNRS, \'Ecole Polytechnique, Institut Polytechnique de Paris, F-75252 Paris, France.}

\author{P. S. Kleij} %\footnote{paula.kleij@gmail.com}}
\affiliation{LSI, CEA/DRF/IRAMIS, CNRS, \'Ecole Polytechnique, Institut Polytechnique de Paris,  F-91120 Palaiseau, France.}
%\affiliation{LULI, Sorbonne Universit\'e, CEA, CNRS, \'Ecole Polytechnique, Institut Polytechnique de Paris, F-75252 Paris, France.}

\author{F. Amiranoff}% \footnote{francois.amiranoff@polytechnique.edu}}
\affiliation{LULI, Sorbonne Universit\'e, CEA, CNRS, \'Ecole Polytechnique, Institut Polytechnique de Paris, F-75252 Paris, France.}

\author{M. Grech}% \footnote{mickael.grech@polytechnique.edu}}
\affiliation{LULI, Sorbonne Universit\'e, CEA, CNRS, \'Ecole Polytechnique, Institut Polytechnique de Paris, F-75252 Paris, France.}

\author{C. Riconda}% \footnote{caterina.riconda@upmc.fr}}
\affiliation{LULI, Sorbonne Universit\'e, CEA, CNRS, \'Ecole Polytechnique, Institut Polytechnique de Paris, F-75252 Paris, France.}

\author{M. Raynaud}
\email{michele.raynaud-brun@polytechnique.edu}
\affiliation{LSI, CEA/DRF/IRAMIS, CNRS, \'Ecole Polytechnique, Institut Polytechnique de Paris,  F-91120 Palaiseau, France.}

%\nodate{\nodate}
\begin{abstract}
Ultra-short high-power lasers can deliver extreme light intensities ($\ge 10^{20}$ W/cm$^2$ and $\leq 30 f$s) and drive large amplitude Surface Plasma Wave (SPW) at over-dense plasma surface. The resulting current of energetic electron has great interest for applications,  potentially scaling with the laser amplitude, provided the  laser-plasma transfer to the accelerated particles mediated by SPW  is still efficient at ultra high intensity. By mean of Particle-in-Cell simulations, we identify the best condition for SPW excitation and show  a strong correlation between the optimum Surface Plasma Wave excitation angle and the laser's angle of incidence that optimize the electron acceleration along the plasma surface. We also discuss how plasma density and plasma surface shape can be adjusted in order to push to higher laser intensity the limit of Surface Plasma Wave excitation. Our results open the way to new experiments on forthcoming multi-petawatt laser systems.
\end{abstract}

\maketitle

\section{Introduction}\label{sec_introduction} 

The interaction of an intense laser pulse with an over-dense plasma, possessing a sharp density gradient, can result in accelerate charged particles with relativistic velocities \citep{wilks92,wilks97,kruer85, brunel87,brunel88,kruer,ren06,beg97}. The irradiation of structured targets \citep{kluge12, jiang14}, such as periodic grooves (gratings) on a metal surface, by ultra short laser pulses are of particular interest for generating intense Surface Plasma Waves (SPWs), that can store the laser energy and efficiently accelerate electrons.

In this scenario, high energy transfer from the laser to the plasma is achieved when the frequency and wavelength of the interacting laser pulse match those given by the SPW's dispersion relation \citep{raether88, kaw70, bigongiari11}. The high intensity and ultra-short laser-plasma interaction regime ($\leq 10^{19}$ W/cm$^2$ and $\leq100 f$s), showed that a significant percentage of electrons trapped in the SPW can be accelerated along the surface in the range of $\sim 10$ MeV~\citep{ceccotti13,riconda15,naseri13,Willingale11,Willingale13,fedeli:16}. High charge electron bunches (up to $\sim 650$ pC) were also observed \citep{fedeli:16, cantono18, raynaud20, zhu20, marini} with applications including the generation of bright sources of ultra-short pulsed X-rays, ultra-fast electron diffraction, tabletop electron accelerators, and ultra-fast electron spectroscopy \cite{Azamoum, liu, tokita,lupetti}. Recently, a scheme exploiting up to date laser techniques was proposed for controlling the duration and amplitude of SPWs by which a laser with an intensity of a few $ 10^{19}$ W/cm$^2$ and a pulse duration of a few tens of $f$s should be able to accelerate electrons up to $\sim 70$MeV \cite{marini}. 
Surprisingly, in these experiments and simulations, the non-relativistic cold dispersion relation successfully defined the conditions of the SPW excitation with laser beam intensity up to $\sim 10^{19}$ W/cm$^2$. 

Extending the regime of ultra-high laser intensity interaction beyond $10^{21}$ W/cm$^2$ can result in surface waves with extremely large amplitudes at the over-dense plasma surface, potentially allowing to obtain unprecedentedly high currents of energetic electrons as well as emitting radiation with interesting characteristics. However, the excitation and survival of these SPWs in the ultra-high laser intensity regime remains an open question, as in this limit the plasma grating can evolve on relatively short time scales, and nonlinear effects can affect the dispersion relation in the relativistic regime. 

In this paper we determine the conditions for improving laser-plasma energy transfer as well as accelerating charged particles by the SPW excitation mechanism in an over-dense plasma with a grating, in the ultra-high laser intensity regime of interaction. We employed 2D Particle-In-Cell (PIC) simulations for laser intensities ranging from $10^{16}$ to $10^{22}$ W/cm$^2$, for various angles of incidence. The influence of both the plasma density and the grating depth of the modulated plasma surface were investigated since  previous studies identified them as important parameters in SPW excitation \citep{fedeli:16, cantono18, raynaud20, zhu20, marini}.

The paper is organized as follows: section \ref{parameter_simul} describes the PIC simulation setup with parameters closely corresponding to recent experiments \citep{fedeli:16, cantono18}. Section \ref{opt_angle} analyses SPW excitation as a function of laser incidence and intensity. 
The results are then compared to analytical values obtained by the dispersion relation for cold SPWs and a heuristic relativistic correction. The importance of considering high density plasma to maintain SPW excitation in the ultra relativistic regime is shown. Section \ref{elec} studies the behavior of accelerated electrons along the plasma surface. A strong correlation is demonstrated between the angle of SPW excitation and the laser's angle of incidence that optimizes electron acceleration along the plasma surface.
The section \ref{depth} investigates the influence of the grating depth at higher laser intensities. Then, in the last section our conclusions are presented.

\section{Parameter of the simulations}\label{parameter_simul}

\begin{figure}[ht]
\begin{center}
\includegraphics[width=4.cm]{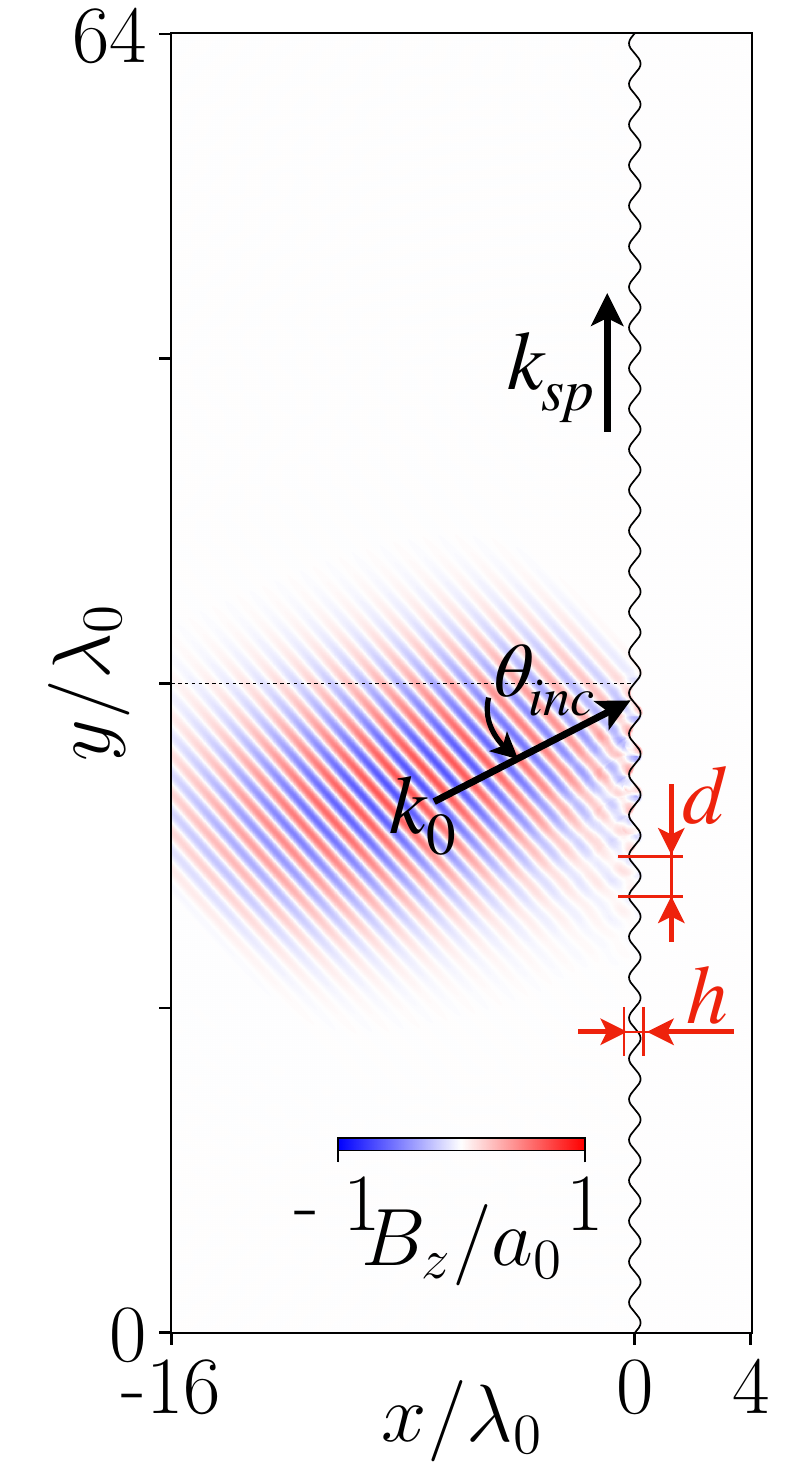}
\caption{Simulation set-up: the laser beam is focused thought an angle $\theta_{inc}$ over the interface of the plasma target with constant electron density $n_0$, grating depth $h$ and period $d$. Here, the red-blue scale represents the magnetic field amplitude of the laser pulse impinging over the target.}
\label{fig:LaserPulseSetup}
\end{center}
\end{figure}

2D3V PIC simulations have been performed with the open-source code SMILEI \citep{smilei}. The geometry is depicted in Fig. \ref{fig:LaserPulseSetup} where the plasma lies in the $(x,y)$ plane for $x\geq 0$, its surface being along the $y$ direction. 

The driven laser is a $P-$polarized Gaussian pulse with a waist equal to $5\lambda_0$ ($=4\mu$m) and a pulse duration equal to $\tau_L=10\lambda_0/c$ ($\simeq27$ fs) full width at half maximum (FHWM), where $c$ is the speed of light in vacuum, and $\lambda_0=0.8\mu m$ is the chosen laser wavelength. The laser pulse impinges the plasma interface through an angle $\theta_ {inc}$ in relation to the normal surface along the $x-$direction. The plasma grating has constant electron density $n_0$ with a sinusoidal-modulated vacuum-plasma interface located at $x_g(y) = (h/2)\,\sin(2\pi y/d)$ where $h$ is the grating depth and $d$ the period. In all cases studied, we considered $d=2\lambda_0$ ($=1.6\mu$m) and we used $h=0.1 \lambda_0$ ($=0.08\mu$m) or $0.4 \lambda_0$ ($=0.32\mu$m) for the grating depth. The plasma consists of electrons with a small initial temperature of $T_e = 50{\rm eV}$ as well as a neutralizing background of ions free to move in the space with initial temperature $T_i/(ZT_e)=0.1$, where $Z=1$ is the atomic number.

In the systematic study we have performed, we selected two values for the plasma density: $n_0=100n_c$ and  $n_0=200 n_c$ where $n_c= \epsilon_0 m_e \omega_0^2/e^2$ ($\omega_0$ is the laser frequency and $\epsilon_0$ the vacuum permittivity). These values are chosen in order to study the theoretical  dependence on the plasma density and are compatible with the plasma density obtained in experiments by ionizing solid gratings  \citep{ceccotti13, fedeli:16, cantono18, zhu20}. Additionally, we varied the laser field strength (normalized vector potential $a_0 \equiv e E_0/(m_e c\,\omega_0)$) from $a_0=0.1$ ($ \sim \times 10^{16}$ W/cm$^2$) to $a_0=50$ ($\sim 4\times 10^{21}$ W/cm$^2$) as may be reached on forthcoming multi-petawatt laser systems, see {\it e.g.} Refs.~\cite{apollon,jeong14}. For any given ($n_0,a_0$), we have performed a parametric scan varying the incidence angle of the laser from typically $\theta_{\rm inc} = 28^{\circ}$ to $\theta_{\rm inc} = 50^{\circ}$ in order to extract the optimal condition for SPW excitation. 

In these simulations, the box extends over $20\lambda_0$ ($=16\mu$m) in the $x$-direction [roughly $16\lambda_0$ ($=12.8\mu$m)  of vacuum and $4\lambda_0$ ($=3.2\mu$m) of plasma], and $64\lambda_0$ ($=51.2\mu$m) in the $y$-direction. The spatial resolution was set to $\Delta x = \Delta y = \lambda_0/128$ ($=0.00625\mu$m). The simulation time step is chosen to be $\Delta t = 0.95~\Delta x/\sqrt{2}$ that corresponds to 95\% of the Courant– Friedrich– Lewy (CFL) condition for the standard finite-difference time-domain (FDTD) solver~\cite{nuter2014}. Every cells contains initially $16$ randomly distributed particles of each species (electrons and ions). Electromagnetic field boundary conditions are injecting/absorbing in $x$ and periodic in $y$. Particle boundary conditions in $x$ are reflecting (left) or thermalizing (right), and periodic in $y$. The simulations were run over until particles or radiation get the position $y=60\lambda_0$ ($=48\mu$m), which determines the final simulation time $t=t_f$. Notice that $t_f$ varies according to the laser incidence angle and it gets larger as $\theta_{inc}$ increases.

\section{Resonance condition for SPW excitation at high intensity}\label{opt_angle}

In order to evidence the condition for SPW excitation as function of the laser intensity, we perform a set of simulations with intensity corresponding to $a_0$ varying from $a_0=0.1$ to $a_0=50$ and incident angle ranging from $\theta_{inc}=28^{\circ}$ to $50^{\circ}$. The plasma grating period and depth are kept constant. Initially, the depth is chosen as $h=0.1\lambda_0$, so that corrections to the dispersion relation due to finite depth are negligible. The SPW dispersion relation in the cold plasma non relativistic limit is \citep{kaw70}: 
\begin{eqnarray}\label{eq:dispRel}
\frac{c^2k^2}{\omega^2} = \frac{\omega_p^2/\omega^2-1}{\omega_p^2/\omega^2-2},
\end{eqnarray}
$k$ and $\omega$ are the SPW wavelength and the frequency, and $\omega_p$ is the plasma frequency. In the presence of high-intensity lasers plasma interaction, and in particular when the laser electric field $E_0$ becomes of the order of $m_e c \omega_0/e$ ({\it i.e.} for a normalized vector potential $a_0 \equiv e E_0/(m_e c\,\omega_0) \gtrsim 1$), it has been proposed \citep{Akhiezer, macchi01,siminos12, raynaud18, macchi18} to correct the response of the electrons by considering an effective electron mass $m_e \rightarrow \gamma_0\,m_e$, with $\gamma_0 \simeq \sqrt{1+a_0^2/2}$ the Lorentz factor of an electron in a plane wave with normalized vector potential $a_0$. In the case of SPW excitation by the laser, we thus consider a heuristic correction to the dispersion relation by replacing $\omega_p^2/\omega^2 \equiv \omega_p^2/\omega_0^2$ by $\omega_p^2/(\gamma_0\,\omega_0^2)$. 
As a consequence, correcting the phase-matching condition leads a $a_0$-dependent optimal angle of incidence for the surface plasma wave excitation:
\begin{eqnarray}\label{eq:optAngle}
\theta_{\rm opt}(a_0) = \arcsin\left( \sqrt{\frac{n_0/(\gamma_0 n_c)-1}{n_0/(\gamma_0 n_c)-2}} - \frac{\lambda_0}{d}\right)\,.
\end{eqnarray}
This results in an optimal angle, $\theta_{opt}$ that increases with the amplitude of the SPW field. For $a_0 \gg 1$ it depends on the parameter $n_0/(\gamma_0 n_c)\sim \sqrt{2} n_0/(a_0 n_c) $. 
In order to verify the validity of this scaling,  we considered two electron densities, $n_0 = 100n_c$ and $200n_c$.

As detailed in the following we find in simulations that at high intensity the resonance is quite broad. Although for  values of $n_0/(a_0 n_c) \lesssim 10$ the correction to the dispersion relation can improve the coupling of the laser with plasma. We notice no further improvement for higher value of $a_0$, and the resonance angle becomes roughly independent of $a_0$. We can then conclude that Eq.~\eqref{eq:optAngle} does not hold at ultra-high intensity. 

To show this let us recall that 
SPW are TM-modes, so their signature can be sought by inspecting the Fourier transform of the $B_z$ component of the magnetic field. Taking into account that the SPW and incident/reflected laser waves have different dispersion relations, filtering in ($k_x,k_y$) Fourier space allows to extract the component associated to the SPW.
Then an inverse Fourier transform is done to obtain the $B_z$ component of the SPW magnetic field in the reconstructed real space domain.

\begin{figure}[ht]
\begin{center}
\includegraphics[width=8.cm]{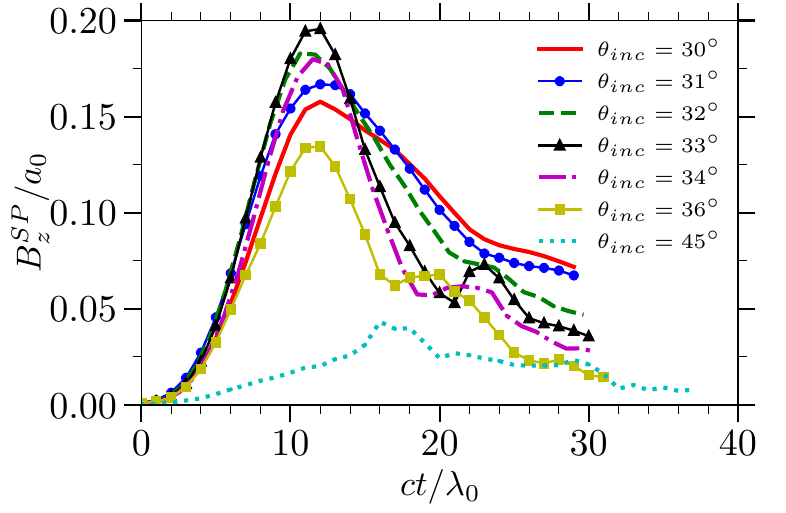}
\caption{SPW $B_z$ field amplitude evolution for $a_0=20$, $n_0=200n_c$ and $h=0.1\lambda_0$, and laser incidence angle in between $30^\circ$ and $45^\circ$, $t=0$ corresponds to the instant of time when the laser pulse reaches the plasma. 
\label{fig:SPWevolution01}}
\end{center}
\end{figure}

The time evolution  of the maximum amplitude of the SPW $B_z$ field normalized to $a_0$ for a typical case ($a_0=20$, $n_0=200n_c$, $h=0.1\lambda_0$ and different values of the laser incidence angle between $30^\circ$ and $45^\circ$) is reproduced in Fig.~\ref{fig:SPWevolution01}. The field component reaches a maximum around $t=12\lambda_0/c$ for an incidence angle of $33^\circ$, named hereafter $\theta_{opt}$ with $t=0$ corresponding to the time when the laser pulse reaches the plasma surface. We notice that the SPW field amplitude does not become larger than the laser field $a_0$, as opposed to what has been found for longer pulses and lower intensities \cite{bigongiari11}. In this short pulse regime ($\simeq27f$s) the SPW excitation does not have time to reach the stationary regime. From the figure we can also see that the resonance condition is not sharp. A laser incident at angles close to the optimal values excite a field with very similar behaviour to the optimal one. This is also due,  as discussed in Ref.~\cite{raynaud20}, to the fact that the width of the incident laser transverse profile induces a spectral mode distribution of the SPW which induces an angular width for the $\theta_{opt}$ equals here to $\sim 4^\circ$. 

\begin{figure}[!ht]
\begin{center}
\includegraphics[width=4.2cm]{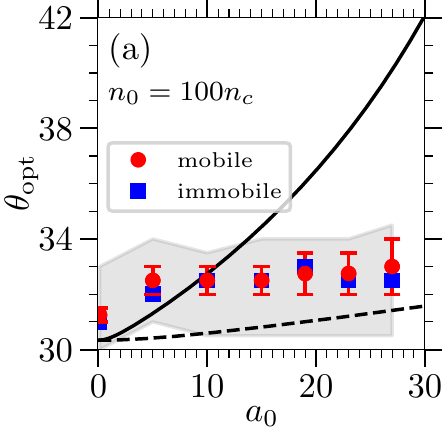} 
\includegraphics[width=4.2cm]{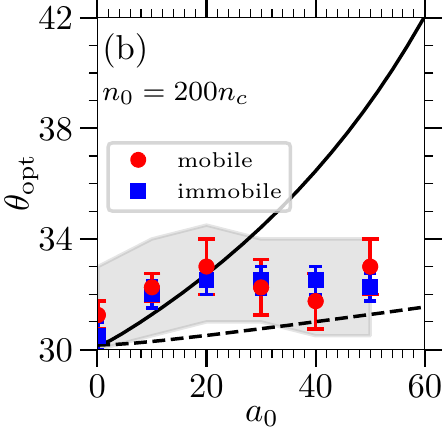} 
\caption{In red (rounds) laser angles of incidence that optimizes the SPW $B_z$ field amplitude as a function of the laser strength parameter $a_0$ for (a) $n_0= 100n_c$, and (b) $n_0=200n_c$. The gray region represents the laser angles of incidence where $\max|B_{z}^{SP}|\gtrsim 0.85 \max|B_{z}^{SP}|$. In blue (squares) we report the results from simulations assuming immobile ions. In both cases, $h=0.1\lambda_0$. The solid (dashed) black line represents the expected value obtained using the dispersion relation for the cold SPW limit with the heuristic relativistic correction as a function of $a_0$ ($a_0/5$) (see the discussion in the text).}
\label{nfig3}
\end{center}
\end{figure}

In Fig.~\ref{nfig3}  we report the optimum laser incidence $\theta_{opt}$ (red dots and error-bar) as a function of $a_0$ for the two plasma densities considered. The $\theta_{opt}$ is obtained by considering for each $a_0$ the angle that corresponds to the peak value of SPW $B_z$ in time (following the same procedure that is illustrated in Fig. \ref{fig:SPWevolution01}). In the panels, the error-bars measure the uncertain measuring the peak value of SPW $B_z$ while the gray shadow identifies the region where $\max|B_{z}^{SP}|\gtrsim 0.85 \max|B_{z}^{SP}|$.
As $a_0$ increases, and in particular for  $a_0 \gtrsim n_0/(10 n_c) $, the incertitude in determining the  optimum angle of the SPW $B_z$ becomes large since many angles correspond more or less to the same maximum value of the field. Moreover, when increasing $a_0$, the normalised amplitude of the field $B^{SP}_z/a_0$ decreases. We notice that going from $a_0\sim 1$ to  $a_0 \sim n_0/(10 n_c) $ results in a reduction of the field amplitude of $\approx45\%$. Further increasing $a_0$ and taking $a_0 \sim n_0/(4 n_c) $ results in a field amplitude reduction of $\approx60\%$ in relation to the field observed when $a_0=1$ (not shown here).

In Fig. \ref{nfig3} we also plot in black the expected value obtained using Eq. (\ref{eq:optAngle}). As anticipated, while at first the values obtained in the simulations fit the equation, for larger values of $a_0$ the resonance angle becomes roughly independent of $a_0$. The threshold, noted $a_{0,T}$ in the following, is about $a_{0,T}=10$ in the case when $n_0=100n_c$ and increases up to $20$ when $n_0=200n_c$ (or, equivalently, $n_0/(a_{0,T} n_c) \sim 10$).  As we can see, even if Eq. (\ref{eq:optAngle}) does not hold, the parameter $n_0/(a_0 n_c)$ is a relevant quantity to describe the laser plasma coupling and the SPW excitation. More importantly, this parameter shows the importance of considering higher density plasma to maintain SPW excitation in the ultra relativistic regime.

In Eq. (\ref{eq:optAngle}) the heuristic correction to the dispersion relation is obtain using the laser parameter $a_0$. In the present simulations the SPW maximum field amplitude is always smaller than $a_0$, and typically, as shown in  Fig. \ref{fig:SPWevolution01}, of the order of $a_0/5$. Therefore, for reference we also report in dashed black line in Fig. \ref{nfig3} the result from Eq.~\eqref{eq:optAngle} considering $a_0/5$ instead of $a_0$ in the $\gamma_0$ function.

\begin{figure}[ht]
\begin{center}
\includegraphics[width=4.2cm]{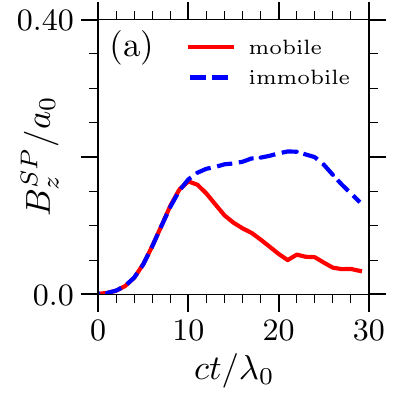}
\includegraphics[width=4.2cm]{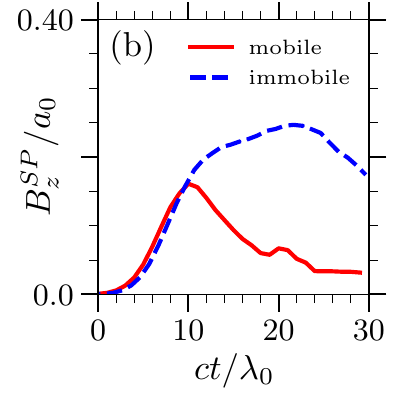}
\caption{SPW $B_z$ field amplitude evolution at $\theta_{inc}=33^{\circ}$ with time for (a) $n=100n_c$, $a_0=27$, and (b) $n=200n_c$, $a_0=50$. $t=0$ corresponds to the instant of time when the laser pulse reaches the plasma. }
\label{nfig4}
\end{center}
\end{figure}

Increasing $a_0$ increases the laser pressure, which may alter the grating and suppress the SPW excitation. To check the importance of this effect and to verify if the relativistic correction  of the dispersion relation (Eq.~\eqref{eq:optAngle}) is recovered, we also performed a set of simulations with immobile ions (represented by blue squares in Fig.~\ref{nfig3} and a blue dashed line in Fig.~\ref{nfig4}). As we can see, the optimal angle is barely modified  when the ions are immobile. However, as shown in  Fig.~\ref{nfig4} where we plot the SPW field amplitude evolution with time for two densities and $a_0>a_{0,T}$, in the case of immobile ions the SPW field survives a longer time and peaks to higher values. This means that the grating deformation affects the SPW field on time scales larger than few laser periods ($\sim 12\lambda_0/c$ here).

Above $a_{0,T}$ the damping of the SPW by the electrons is large, resulting in strong  electron acceleration along the surface trapped in the SPW \citep{riconda15,fedeli:16, raynaud20}. In the next section of this paper we consider the SPW evolution as related to the electron dynamics along the grating.
\section{Electron acceleration along the plasma surface}\label{elec}

\begin{figure}[ht]
\begin{center}
\includegraphics[width=4.2cm]{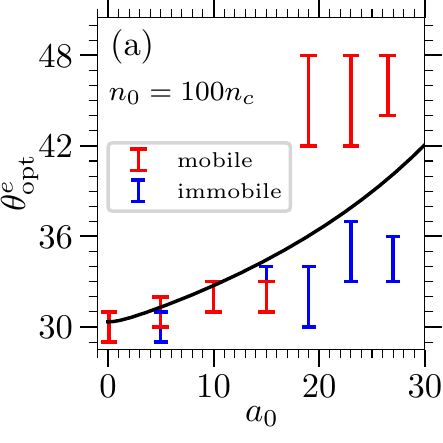}
\includegraphics[width=4.2cm]{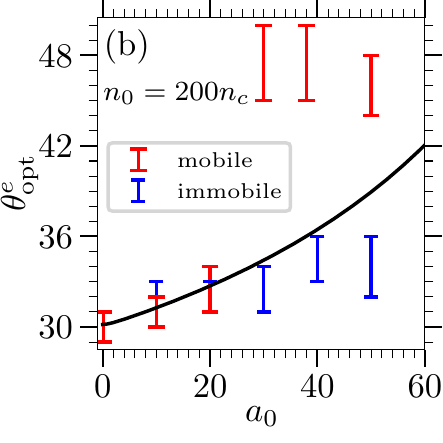}
\caption{In red (bars), angle of incidence of the laser that optimize electron bunches energy propagating along the plasma surface ($\theta_{opt}^e$) as a function of the laser strength parameter $a_0$ for (a) $n_0= 100n_c$, and (b) $n_0=200n_c$. In blue (bars) is reported results from simulations assuming immobile ions. In both cases, $h=0.1\lambda_0$. The solid black line reports the optimal angle of SPW excitation obtained using the dispersion relation for cold SPW with the heuristic relativistic correction (see the discussion in the text).}
\label{fig3new}
\end{center}
\end{figure}

As mentioned in the introduction, SPW excitation resulting from high intensity ultra-short laser plasma interaction ($\leq 10^{19}$ W/cm$^2$ and $\leq100 f$s) has been shown to be an efficient way to increase the acceleration of high charge electron bunches along the plasma surface up to $\sim 10$MeV and $\sim 650$pC \citep{cantono18,riconda15,naseri13,Willingale11,marini,Willingale13,fedeli:16,raynaud20,zhu20}. 
Using the same laser intensities and plasma densities as in the previous section, we will first analyze the maximum energy of the electrons that propagate along the plasma surface as a function of the laser angle of incidence.
The results are summarized in Fig. \ref{fig3new} where we report the optimal laser's angle of incidence, $\theta^e_{opt}$ (which optimizes the formation of high energetic electron bunches propagating along the plasma surface) as a function of the laser strength parameter $a_0$ for (a) $n_0= 100n_c$, and (b) $n_0=200n_c$ (case $h=0.1\lambda_0$). To identify the electrons that propagate along the surface, we have defined the emission angle $\phi_e=\tan^{-1}(p_y/p_x)$ and selected electrons with $\phi_e=90^{\circ}\pm 3^{\circ}$. The bars indicates the range of angles of the laser incidence giving the highest electron energy. This was determined by analyzing for each angle the energy spectrum of the electrons propagating along ($\phi_e=90^{\circ}\pm 3^{\circ}$) the plasma surface. Notice that for the angles considered in the error bar the electron peak energy is about the same within a percentage of up to $10\%$.

As before, we have considered both mobile and immobile ions with the same color code as in Fig.~\ref{nfig3} (red - mobile, blue -immobile). Comparing Fig.~\ref{nfig3} and Fig.~\ref{fig3new} we find at low laser intensity a strong correlation between the optimum angle of SPW excitation and the laser angle of incidence that optimize the electron acceleration along the plasma surface.
The optimum angle giving the highest energy of the electron bunch propagating along the surface is $\sim 31^{\circ}$ for $a_0\sim 1$ and increases slightly up to $\sim 33^{\circ}$ with $a_0$ until it reach $a_{0,T}$. It confirms the robustness of the SPW excitation in this range of intensity.
 
Above $a_{0,T}$, we observe for the realistic simulations (mobile ions) that the laser incidence angle that optimize the electron bunch propagating along the surface is no longer the same one that optimize the SPW field.
The transition occurs for $a_0$ around $20$ if the plasma density is $n_0 = 100n_c$, and around $30$ if $n_0 = 200n_c$. However when considering simulations with immobile ions (blue bars) we recover the result of the previous Fig.~\ref{nfig3}: the optimal angle for electron acceleration coincides with the optimal angle for SPW excitation. This shows that the electrons dynamic is sensitive to the grating deformation. 
Other acceleration mechanisms along and across the surface have been suggested associated to the laser absorption \citep{Macchi19} : indeed we find that above $a_{0,T}$ acceleration by SPW is not the main mechanism of electron acceleration, and the fast electrons angular distribution is much wider.  This will be  discussed in more detail in the next section, but we can anticipate that the analysis of the electron phase space confirms this hypothesis, since above $a_{0,T}$ the electron velocity distribution does not show the characteristic behaviour of the acceleration by SPW, namely bunches with periodicity equal to the SPW wavelength, and directed along the surface \cite{raynaud04}. 
Finally, we checked the effect of the laser on the plasma surface examining the spatial ion density distribution in two different time scales. Both an increase in plasma density due to the radiation pressure and an expansion of the plasma is observed (not reported here). In the short time interval (comparable to the laser pulse duration), the diffraction grating is distorted and the plasma is pushed, which results in a large increase in the local plasma density. In the second and long scale that happens few cycles after the laser-plasma interaction, the plasma expansion creates an under-dense region in front of the target. That also might have a major effect on the laser absorption mechanism and to define the optimal angle to the electron acceleration. The effect of under-dense sheet in front of the plasma surface has been investigated in Ref. \citep{zhu20}.

To overcome the possible limitation of SPW-laser coupling at high laser intensity, we now consider the influence of the target grating depth that, when chosen appropriately, can significantly improve the acceleration by SPW.

\section{Recovery of SPW acceleration by adapting the grating depth}\label{depth}
 
In laser-solid interaction and also at high laser intensity where plasma is created, it is well known that the ratio between the target grating depth and the grating periodicity plays a major role in the SPW excitation \cite{Hutley, raynaud18}.
Thus here, in order to find the optimum grating parameters for SPW excitation in the ultra high laser intensity regime ($a_0\geq 25$), we have redone the PIC simulations increasing the grating depth of the plasma to $h=0.4\lambda_0$.

\begin{figure}[ht]
\begin{center}
\includegraphics[width=4.2cm]{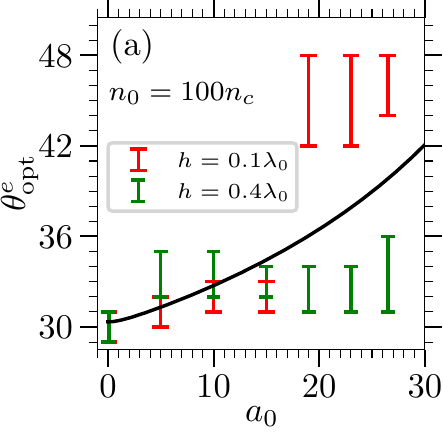}
\includegraphics[width=4.2cm]{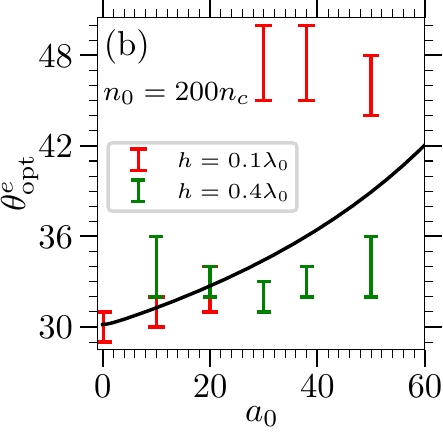}
\caption{Optimal angle of incidence of the laser that optimize electron bunches energy propagating along the plasma surface ($\theta_{opt}^e$) as a function of the laser strength parameter $a_0$ for (a) $n_0= 100n_c$, and (b) $n_0=200n_c$ (case $h=0.1\lambda_0$ in red and $h=0.4\lambda_0$ in green). The black line reports the expected value obtained using the dispersion relation for cold SPW with the heuristic relativistic correction (see the discussion in the text).}\label{fig7new}
\end{center}
\end{figure}

\begin{figure}[ht]
\begin{center}
\includegraphics[width=8.cm]{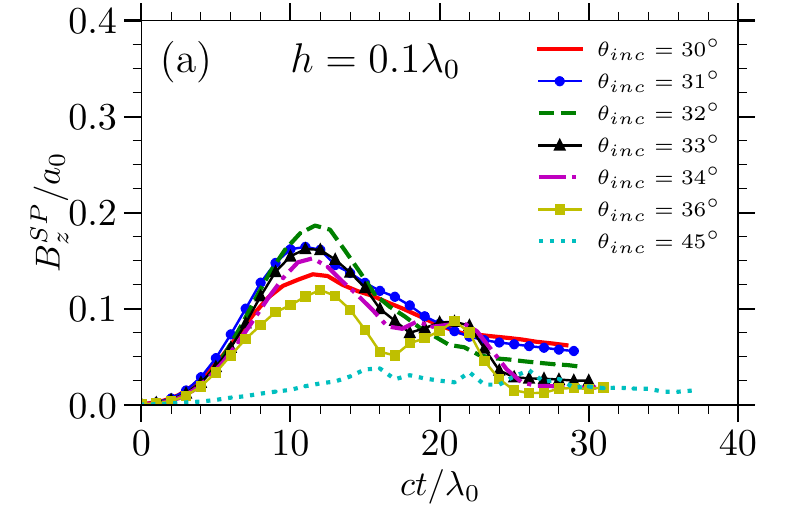}
\includegraphics[width=8.cm]{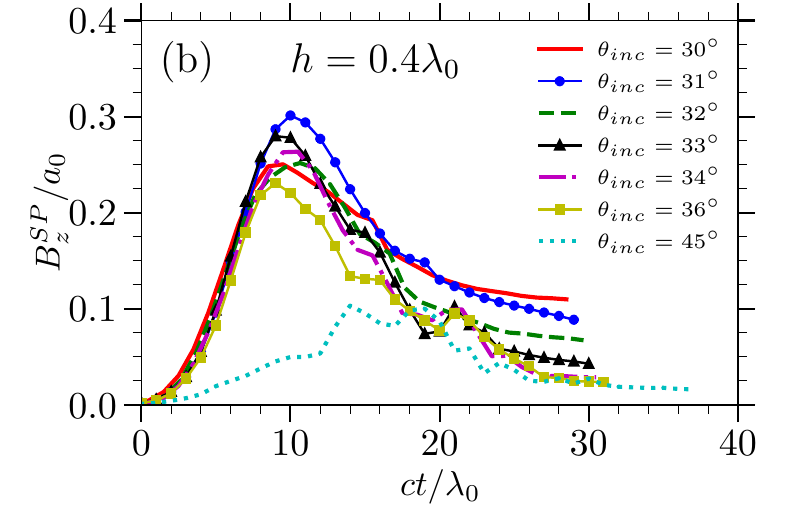}
\caption{SPW $B_z$ field amplitude evolution with time for $a_0=30$, $n_0=200n_c$, laser incidence angle in between $30^\circ$ and $45^\circ$, and $h=0.1\lambda_0$ (a) and $h=0.4\lambda_0$ (b). $t=0$ corresponds to the instant of time when the laser pulse reaches the plasma.}
\label{fig:SPWevolution}
\end{center}
\end{figure}

In Fig. \ref{fig7new} we compare the optimal angle of incidence of the laser that optimize electron bunches energy propagating along the plasma surface (bars) found in the previous section for $h=0.1\lambda_0$ (in red) with the one found for $h=0.4\lambda_0$ (in green) keeping unchanged the other parameters. As we can see in the case $h=0.4\lambda_0$ the optimum angle for particle acceleration remains between $30^{\circ}$ and $36^{\circ}$ and coincides with the optimum angle for SPW excitation as presented in Fig. 3. As in section III the best laser incidence angle to excite highly energetic electron bunches stay roughly constant and does not scale with the laser strength. This is illustrated as an example by the simulations at $a_0=30$. In Fig. \ref{fig:SPWevolution} we plot the maximum $B_z$ field amplitude evolution in time for different values of laser incidence angle and $h=0.1\lambda_0$ (a) and $h=0.4\lambda_0$ (b). Comparing Fig. \ref{fig:SPWevolution}a, where $a_0=30$ and $h=0.1\lambda_0$, the time evolution of the field is quite similar to that observed in Fig. \ref{fig:SPWevolution01} where $a_0=20$ and $h=0.1\lambda_0$. However, when we increase the grating's depth, the value of the field amplitude is larger and the optimal angles ($31^{\circ}-33^{\circ}$) coincide with the optimal angles for electron acceleration in Fig. \ref{fig7new}b. As a consequence with the deeper grating we expect both that the electrons are mainly accelerated by the SPW and that the maximum energy gained by the electrons is higher than if the grating is shallow.

\begin{figure}[ht]
\begin{center}
\includegraphics[width=8.cm]{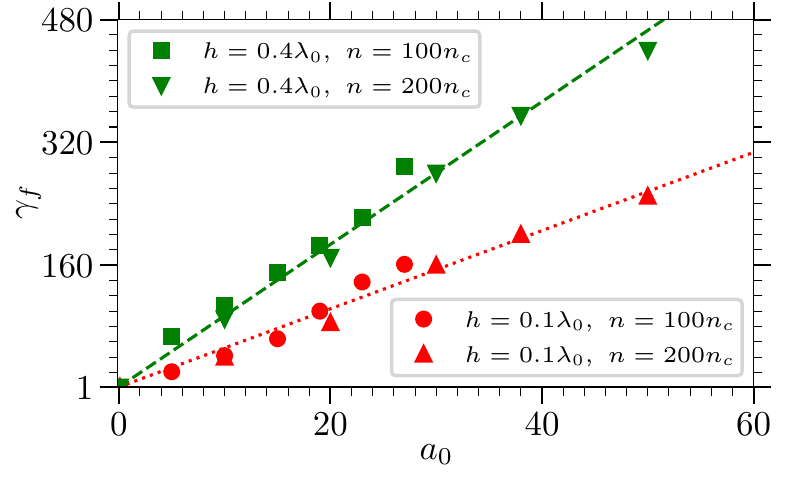}
\caption{Maximum value of gamma factor, $\gamma_f$ along the target's surface, observed at the end of simulation as a function of a function of the laser strength parameter $a_0$ for $n_0= 100n_c$, and $n_0=200n_c$ (case $h=0.1\lambda_0$ in red and $h=0.4\lambda_0$ in green). The dashed lines represents the general tendency of the results. }\label{figspec}
\end{center}
\end{figure}

In  Fig. \ref{figspec}  we show the maximum value of the gamma factor, $\gamma_f$ along the target's surface, for the electrons observed at the end of simulation as a function of the laser strength parameter $a_0$, taking $\theta_{inc}=\theta^e_{opt}$ and the parameters used in the Fig. \ref{fig7new}. As expected we observe that the energy transfer is better when the gratings are deeper ($h = 0.4 \lambda_0$) than when they are shallow ($h =0.1 \lambda_0$) in the high-intense regime.
The red dotted line is the function $\gamma_f=1+5.1a_0$ that fits the data when $h=0.1\lambda_0$ and the green dashed curve is the function $\gamma_f=1+9.3a_0$ that fits the data when $h=0.4\lambda_0$.

\begin{figure}[ht!]
\begin{center}
\includegraphics[width=3.9cm]{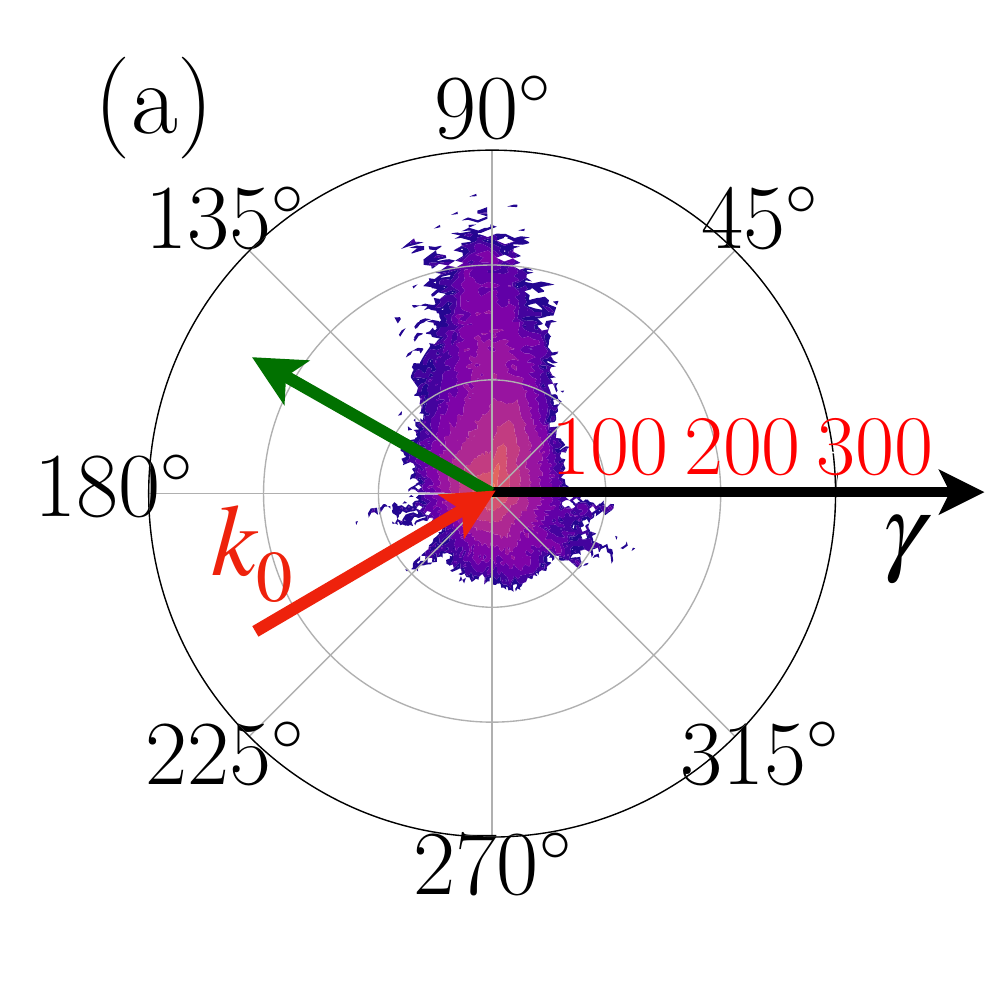}
\includegraphics[width=3.69cm]{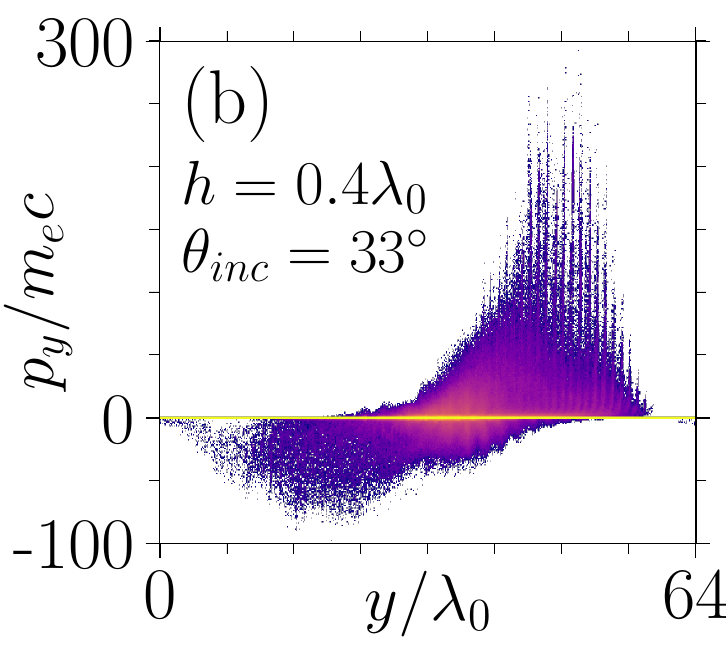}\\
\includegraphics[width=3.9cm]{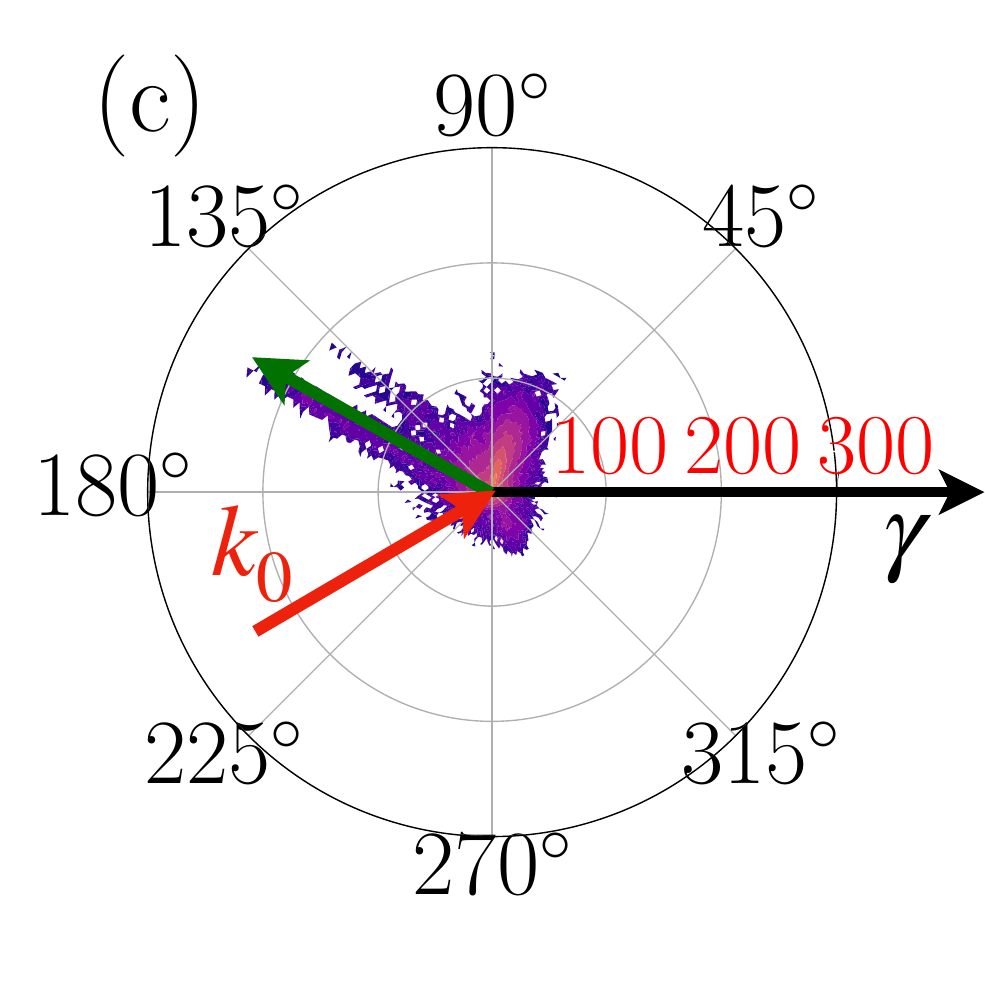}
\includegraphics[width=3.69cm]{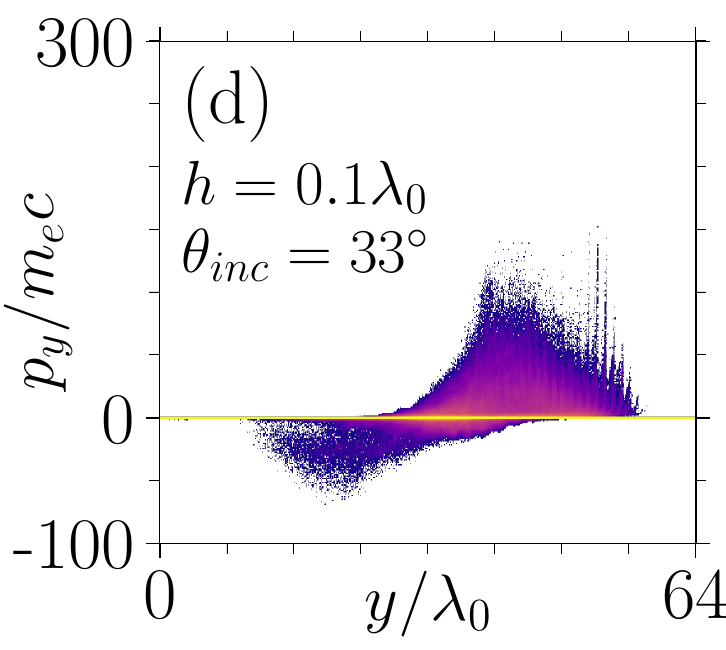}\\
\includegraphics[width=3.9cm]{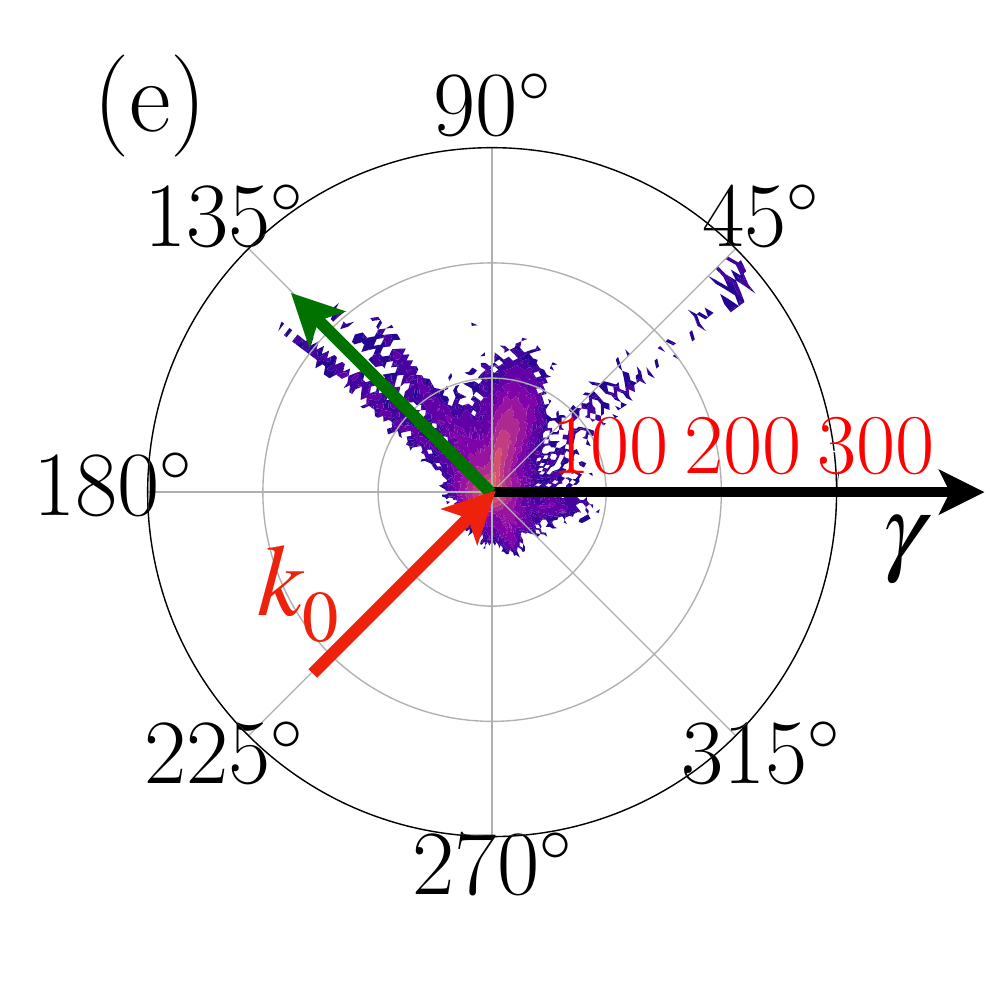}
\includegraphics[width=3.69cm]{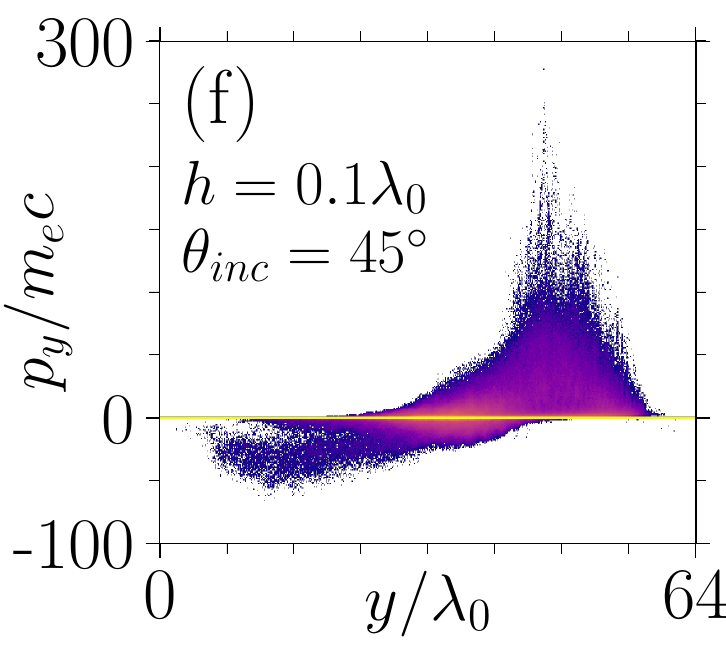}\\
\vspace{0.5cm}
\includegraphics[width=8 cm]{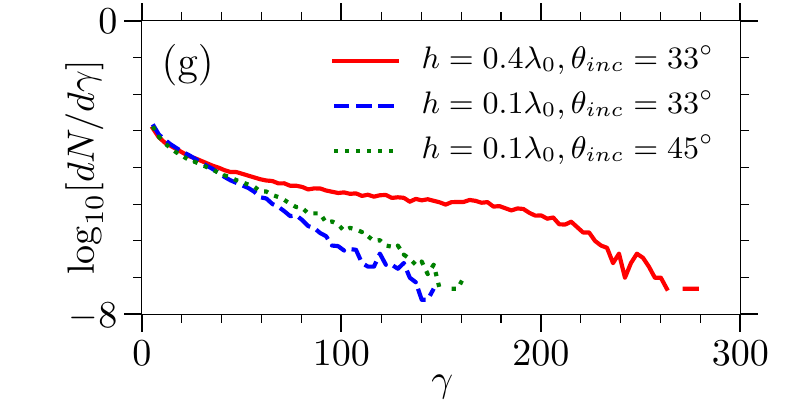}\\
\caption{For $\theta_{inc}=33^{\circ}$, $a_0= 30$, $n_0=200n_c$ and $h=0.4\lambda_0$, (a) electron energy distribution at $t=t_f$. The plasma surface is along $90^{\circ}$, the red arrow shows the direction of the incident laser beam and the green arrow the reflected one; (b) phase space ($p_y/m_e c$,$y/\lambda_0$) of the electrons in the simulation box; the panels (c) and(d) [(e) and (f)] represent the same as the panels (a) and (b) for $h=0.1\lambda_0$ and $\theta_{inc}=33^{\circ}$ [$\theta_{inc}=45^{\circ}$]; 
(g) spectrum of the electron bunches along the surface for the tree parameter sets discussed.}
\label{fig:particleacceleration2}
\end{center}
\end{figure}

A more detailed analysis of the electron dynamics can be inferred from their energy distributions as a function of the propagation angle and from their phase space ($p_y/m_e c$,$y/\lambda_0$). If $h=0.4\lambda_0$ and  $\theta_{inc}=33^{\circ}$, a large amount of highly energetic electrons propagates along the surface $\phi_e=90^{\circ}$ (Fig. \ref{fig:particleacceleration2}(a)), and the phase space shows bunches distanced by a wavelength  (Fig.~\ref{fig:particleacceleration2}(b)), consistent with the SPW acceleration mechanism.\\ %

This is very different from the case with $h=0.1\lambda_0$, and $\theta_{inc}=33^{\circ}$ reported in Fig. \ref{fig:particleacceleration2} (c) and (d) or $h=0.1\lambda_0$, and $\theta_{inc}=45^{\circ}$ reported in Fig. \ref{fig:particleacceleration2} (e) and (f). We observe for these lasts two parameters sets that the faster electrons are accelerated mainly along the direction of the incident and reflected laser beam and fewer electron are found propagating along the surface at $90^{\circ}$. Moreover a large amount of fast electrons are pushed inside the plasma. It is worth to point out that although the peak energy is reduced in this configuration, the laser plasma coupling is still large so that this configuration might be a way to enhance TNSA at the rear of the thin target \citep{heron20}.
In such a limit, the SPW field when present (Fig. \ref{fig:particleacceleration2} (c) and (d)) is weak and the SPW wave is no longer the predominant acceleration mechanism. This might be attributed to the grating deformation due to laser pressure which prevents laser-SPW coupling.

We can thus conclude that a deeper grating allows to recover the exciting of SPW in the ultra high intensity laser regime and acceleration along a preferential direction. This effect is evident in Fig. \ref{fig:particleacceleration2}(g) when comparing the electron's spectra (selecting only the ones emitted parallel to the target $\phi_e=90^{\circ}\pm 6^{\circ}$) for  $h=0.1\lambda_0$ (in blue) and $h=0.4\lambda_0$ (in red), with $\theta_{inc}=33^{\circ}$ in both cases. The electron energy obtained when increasing the grating depth is increased by a factor of two for the deepest grating and the optimal angle.
Instead for $h=0.1\lambda_0$ the energy spectrum changes very little between $\theta_{inc}=33^{\circ}$ and $45^{\circ}$ (in green), even if, when comparing the phase space $p_y/m_e c$,$y/\lambda_0$ for both incident angles (Fig.~\ref{fig:particleacceleration2} (d) and (f)) we observe a small signature of the SPW excitation (bunching of the phase space), that is lost at $45^{\circ}$. 

\begin{figure}[ht!]
\begin{center}
\includegraphics[width=3.9cm]{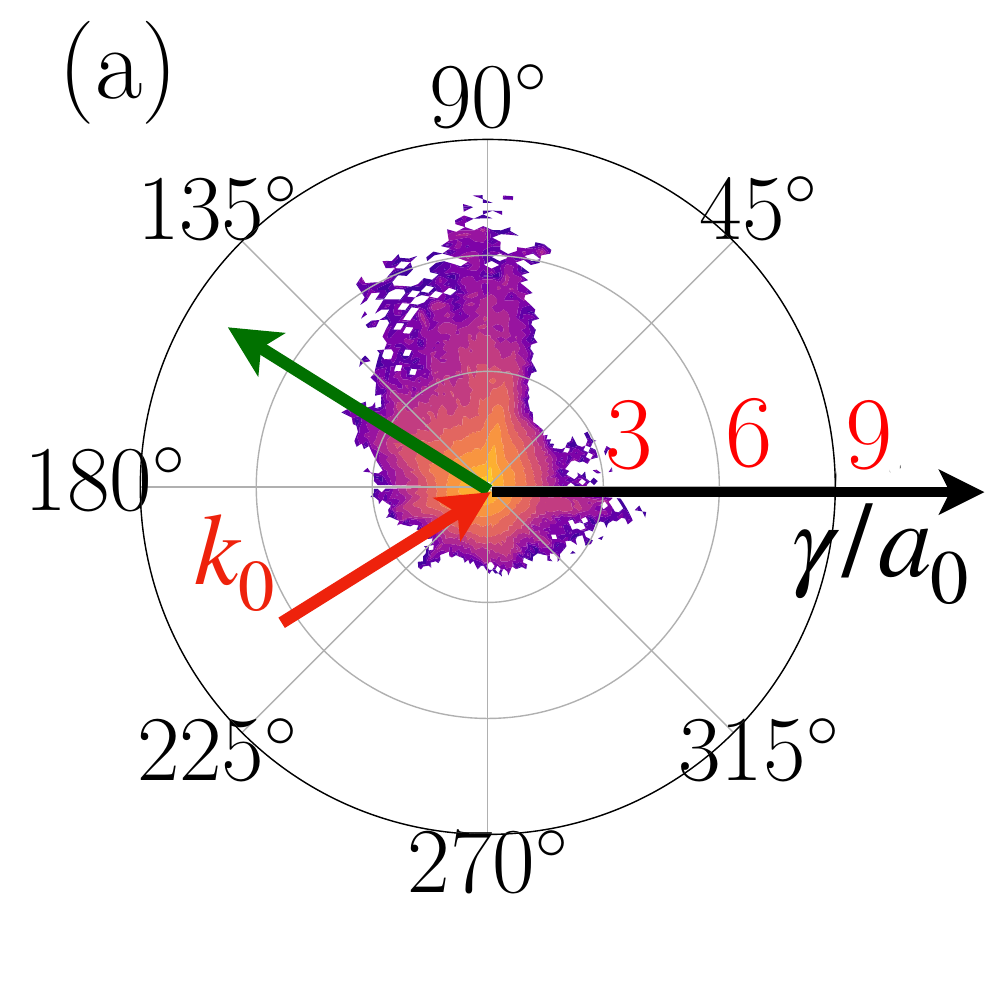}
\includegraphics[width=3.9cm]{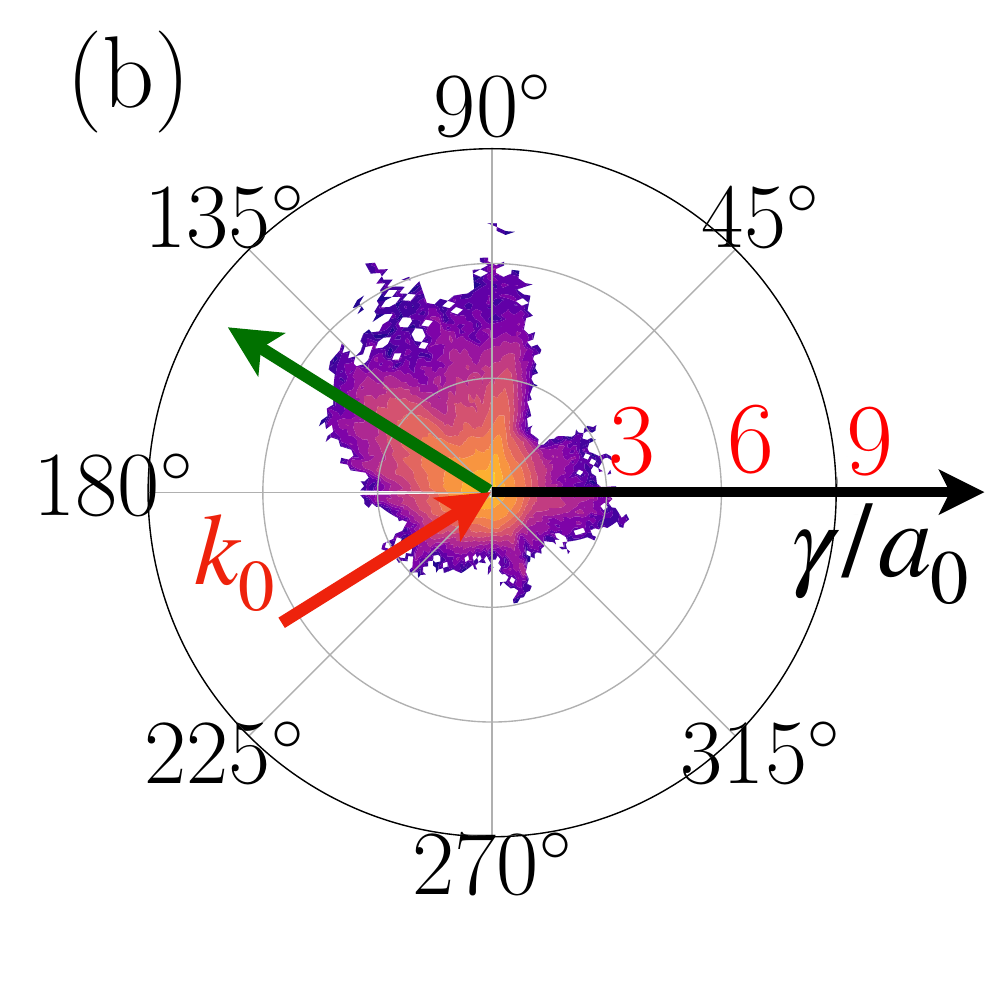}\\
\caption{Electron energy distribution at $t=t_f$, $\theta_{inc}=33^{\circ}$ and $h=0.4\lambda_0$ and, $a_0=100$ (a) and $a_0=200$ (b). The plasma surface is along $90^{\circ}$, the red arrow indicates the direction of the incident laser beam, and the green arrow, the reflected one. Note that although the ratio $\gamma/a_0$ is about the same in both panels, $\gamma_f$ is about $800$ in (a) and $1600$ in (b).} 
\label{fig:temp}
\end{center}
\end{figure}

To conclude this section we verified that for  $h=0.4\lambda_0$, the SPW is still excited even at significantly higher laser intensities. In Fig.~\ref{fig:temp} the electrons emission spectrum assuming two extreme laser conditions (a) $a_0=100$, and (b) $200$ is shown. There, the plasma density is equal to $n_0=200n_c$. From the panels, we observe a large increase of the electron energy achieving $\gamma_f/a_0\sim 7, 8$ ($\gamma_f\approx 800$ for $a_0=100$ and $\gamma_f\approx1600$ for $a_0=200$), even if for the largest laser strength $a_0=200$ (Fig.~\ref{fig:temp} (b)), the angular distribution of the electrons tends to increase. Our results show that, even in the very high-intensity regime of interaction, there is good evidence that SPW excitation and the consequent electron acceleration still present when the diffraction grating is correctly chosen. However, they do not account for additional processes that may set at extreme intensities, such as radiation reaction or quantum effects (like pair creation)\cite{niel18}. These processes are under investigation and remain beyond the scope of this work.

\section{Conclusion}\label{sec_conclusions}
In this work, we consider a laser pulse impinging on an over dense plasma, whose surface presents a periodic modulation (grating), in order to generate large amplitude Surface Plasma Waves (SPWs). Key parameters were obtained for optimising laser-plasma coupling in the ultra-relativistic regime ($\sim10^{22}$ W/cm$^2$).  
A systematic study in function of the laser incidence angle and intensity, $a_0$, employing the SMILEI Particle-In-Cell simulations, showed that at ultra high laser intensities ($a_0\ge30)$ the SPW resonance angle becomes roughly independent of $a_0$. A strong correlation was also observed between the optimum SPW excitation angle and the laser's angle of incidence that optimizes electron acceleration along the plasma surface. The production of high energetic electron bunches is analysed as well as the appropriate values of plasma density and surface shape to ensure SPW survival at ultra-high laser intensity. 
Furthermore, the parameter $n_0/(a_0 n_c)$ is shown as crucial for describing laser plasma coupling and SPW excitation, as it highlights the importance of the prior consideration of higher density plasma to maintain SPW excitation in the ultra relativistic regime. Finally, as high-intense lasers illuminating the grating inevitably distorts it, increasing the grating's depth provides a more robust condition for SPW excitation.
This may be a way to obtain unprecedentedly high currents of energetic electrons as well as emitting radiation with interesting characteristics thereby paving the way to new experiments on forthcoming multi-petawatt laser systems.

\vspace{1cm}

\section*{Acknowledgement}
P.S.K. was supported by the CEA NUMERICS program, which has received funding from the European Union's Horizon 2020 research and innovation program under the Marie Sklodowska-Curie grant agreement No. 800945.
Financial support from Grant No. ANR-11-IDEX-0004-02 Plas@Par is acknowledged. Simulations were performed on the Irene-SKL machine hosted at TGCC- France, using High Performance Computing resources from GENCI-TGCC (Grant No. 2018-x2016057678). We acknowledge PRACE for awarding us access to Irene-SKL.
Technical support from the SMILEI dev-team was also greatly appreciated.

\end{document}